\documentclass[a4paper,12pt]{article}   	
\usepackage{geometry}                		
\usepackage{graphicx}				
\usepackage{amssymb}
\usepackage{amssymb,amsthm,amsmath}
\usepackage{url,hyperref,lineno,microtype,subcaption}

\title{\Huge { Covariant description of the colloidal dynamics on curved manifolds}}

\begin{document}

\def\Authors{\begin{center}{\large Pavel Castro-Villarreal$^{a, 1}$, C\'esar O. Solano-Cabrera$^{b,2}$ and Ram\'on Casta\~neda-Priego$^{b, 3}$}\end{center}}

\def\Address{a. Facultad de Ciencias en F\'isica y Matem\'aticas,
Universidad Aut\'onoma de Chiapas, Carretera Emiliano Zapata, Km. 8, Rancho San Francisco, 29050 Tuxtla Guti\'errez, Chiapas, M\'exico\\
b. Divisi\'on de Ciencias e Ingenier\'{\i}as, Universidad de Guanajuato, Loma del Bosque 103, 37150, Le\'on, Guanajuato, M\'exico.
}

\def\corrAuthor{Corresponding Authors}

\def\corrEmail{ {pcastrov@unach.mx}$^{1}$,\\{o.solanocabrera@ugto.mx}$^{2}$\\ {ramoncp@fisica.ugto.mx}$^{3}$
}

\date{}	
\maketitle

\Authors 

\Address 


\begin{abstract}
Brownian motion is a universal characteristic of colloidal particles embedded in a host medium, and it is the fingerprint of molecular transport or diffusion, a generic feature of relevance not only in Physics but also in several branches of Science and Engineering. Since its discovery, Brownian motion or colloid dynamics has been important in elucidating the connection between the molecular details of the diffusing macromolecule and the macroscopic information of the host medium. However, colloid dynamics is far from being completely understood. For example, the diffusion of non-spherical colloids and the effects of geometry on the dynamics of either passive or active colloids are a few representative cases that are part of the current challenges in Soft Matter Physics. In this contribution, we take a step forward to introduce a covariant description of the colloid dynamics in curved spaces. This formalism will allow us to understand several phenomena, for instance, the effects of curvature on the kinetics during spinodal decomposition and the thermodynamic properties of the colloidal dispersion, just to mention a few examples. This theoretical framework will also serve as the starting point to highlight the role of geometry on colloid dynamics, an aspect that is of paramount importance to understanding more complex phenomena, such as the diffusive mechanisms of proteins embedded in cell membranes.

{\scriptsize
{ \section*{\small Keywords:} Diffusion, Brownian motion, Colloids, Smoluchowski equation, curved manifold}} 
\end{abstract}
\corrEmail

\section{\LARGE Introduction}
Since the pioneering work of Einstein \cite{Einstein}, Brownian motion has become the paradigm for the description and understanding of a large variety of diffusion processes that are present in a large variety of physical, biological, and chemical systems. In recent years, the dynamics of macromolecules and nano-particles on surfaces or curved spaces has been the subject of intensive investigations, mainly due to the particle diffusion shows a richer dynamical behavior at different time scales \cite{Apaza2018,villada2021} than its counterpart in open and flat geometries; it can be either subdiffusive ($\alpha<1$) or superdiffusive ($\alpha>1$), i.e., the mean-square displacement $\left<x^{2}\left(t\right)\right>$ does not increase strictly linearly in time, but it behaves as $\propto t^{\alpha}$. Particularly, diffusion plays a key role in the dynamics of molecular motors moving along heterogeneous substrates \cite{kafri2005}, in the transport of biomacromolecules in the cell due to crowding \cite{Basak2019,ando2010}, and in the lateral diffusion of proteins on fluctuating membranes \cite{Ramadurai2009,ALENGHAT2013}, just to mention a few examples. Most of the diffusion properties on surfaces depend strongly on the generic features of the surface or, strictly speaking, on the surface geometry \cite{tarjus2010}. Of course,  the particle dynamics is not only influenced by geometrical constrictions but also local and thermodynamic properties experience the effects of the manifold where the particles are embedded \cite{Ramirez2017,Quintana2018,D0CP06474B}.

A great effort for understanding Brownian motion on surfaces can be found in colloidal soft matter, where the dynamics of colloidal particles on quasi-two-dimensional geometries have been both experimentally and theoretically investigated by using optical techniques, like digital videomicroscopy, computer simulations, and theoretical approximations \cite{Sarmiento2016,Villanueva2019}. Nonetheless, such investigations deal basically with (almost) flat surfaces, i.e., without including curvature effects. The interest in the use of colloids resides in the fact that they are tiny (nanometer- to micrometer-sized) particles and typically are considered model systems because, among other interesting features \cite{Castaneda2021}, their characteristic time and length scales are experimentally accessible, which allow us to follow the colloidal dynamics and transport processes in real-time \cite{Castaneda2021}. Furthermore, since the colloidal interactions are relatively weak, colloids are highly susceptible to external forces, and hence their static and dynamical properties can be controlled through the application of external fields or by imposing geometrical restrictions, see, e.g., Ref. \cite{Castaneda2021} and references therein. Then, colloids represent an ideal model system to account for the effects of geometry on the nature and dynamics of many-body systems. 

In particular, it has already been demonstrated, and experimentally corroborated, that the curvature dependence of a fluctuating membrane affects the diffusion processes of molecules on the membrane surface \cite{Holyst,Faraudo02,Zhong2017}. These geometrical effects, although important, are still difficult to interpret. The lack of a precise interpretation resides in the fact that, unfortunately, there is no formal or unique way to define diffusion on a curved surface (see, for instance, \cite{Holyst,Faraudo02}). In fact, the description of the colloid dynamics in curved spaces is a non-trivial task; it represents a formidable physical and mathematical challenge. Recently, one of us proposed the generalization of the Smoluchowski equation on curved spaces \cite{Castro2010}. Furthermore, Castro-Villarreal also put forward different geometrical observables to quantify the displacement of a single colloidal particle \cite{Castro2014b}. Within this approach, it was shown that the geodesic mean-square displacement captures the intrinsic elements of the manifold, whereas the Euclidean displacement provides extrinsic information from the surface. An interesting extension of the theory now provides the description of the motion of active Brownian particles \cite{Castro2018}, where the mean-square geodesic displacement captures the relationship between the curvature and the activity of the active colloid. This theoretical framework provided evidence that an active Brownian particle experiences a dynamical transition in any compact surface from a monotonic to an oscillating behavior observed in the mean squared geodesic displacement \cite{Castro2018}; a theoretical prediction of a dynamic transition of this type can be established using a {\it run-and-tumble} active particle confined on a circle $S^{1}$. Recently, this prediction was corroborated in experiments using a nonvibrating magnetic granular system, see, e.g., Ref. \cite{Castro2023}. However, we still face challenges in colloid dynamics on curved manifolds, for example, the generalization of this approach to the situation where the colloids interact not only with other macromolecules, i.e., direct forces, but also the inclusion of all those geometrical mechanisms originating from the curvature.

The aforementioned theoretical formalism has also allowed us to determine the equation of motion of interacting colloids in curved spaces; a generalized Ermack-McCammon algorithm has been developed to study a broader class of transport phenomena in curved manifolds \cite{Castro2014}. 
Interestingly, the predictions of the particle transport in non-Euclidean spaces have been partially corroborated in a series of experiments \cite{villada2021,Zhong2017}; superparamagnetic colloids embedded in a circle and subjected to external magnetic fields \cite{villada2021} and polystyrene nanoparticles diffusing on highly curved water-silicone oil interfaces \cite{Zhong2017}. However, further experimental, computational, and theoretical studies are needed to better understand the rich diffusion mechanisms, particle distribution, and  thermodynamic properties that emerge in colloidal dispersions when the curvature of the space plays an important role. To this end, we propose, as a starting point, the covariance description of the colloid dynamics, which is explicitly explained in the next section.

\section{\LARGE Covariance description of the colloid dynamics} 

As discussed above, one of the main challenges to understanding the effects of geometry on the dynamics of colloids embedded in a curved space is to develop experimental tools and theoretical frameworks that account for the transport properties that occur on the manifold. Below, we then provide the first preliminary steps to build a covariant theoretical formulation of the dynamics of an interacting colloidal system based on the many-body Langevin equation in the so-called overdamped limit \cite{Castro2014}, which allows us to deduce a Smoluchowski equation  \cite{Dhont1996} for the interacting system on the manifold. Before starting with the covariant formulation, let us introduce our notation. Let us consider the colloidal system confined on a $d-$dimensional manifold  $\mathbb{M}$ embedded in a $d+1-$dimensional Euclidean space $\mathbb{R}^{d+1}$ described with the parameterization ${\bf X}: U\subset \mathbb{R}^{d} \rightarrow \mathbb{R}^{d+1}$, where  a particular point in $\mathbb{M}$ is given by ${\bf X}(x)$, being $x\equiv\left(x^{1}, x^{2}, \cdots, x^{d}\right)\in U$ local coordinates of the neighborhood $U$. Using the embedding functions ${\bf X}(x)$, one can define a Riemannian metric tensor by $g_{\alpha\beta}={\bf e}_{\alpha}\cdot{\bf e}_{\beta}$, where ${\bf e}_{\alpha}=\frac{\partial}{\partial x^{\alpha}}{\bf X}\left(x\right)$, with $\alpha=1, \cdots, d$.  Further notions like normal vector, extrinsic curvature tensor, and Weingarten-Gauss equations are introduced in Appendix A from \cite{Castro2014b}.   Typically, spatial dimensions of interest are $d=1$ and $d=2$.

As we have pointed out above, our starting point to describe the dynamics of colloids confined in a curved manifold is based on a previous contribution \cite{Castro2014}, where it is posed the many-body Langevin stochastic equations in the overdamped regime, {\it i.e.}, the  diffusive time scale, in local coordinates,
\begin{equation}
\dot{x}_i^\alpha=
\frac{1}{\zeta}{\bf e}^{\alpha}\left(x_{i}\right)\cdot \left [{{\bf f}}_{i}\left(t\right)+\sum_{i\neq j}{\bf F}_{ij}\left(x_{i}, x_{j}\right)\right ],
\label{many-particle-langevin}
\end{equation}
where $\zeta$ is the friction coefficient, and with $x_{i}^{\alpha}$ being the $i-$th particle position with $i=1, \cdots, N$ and $\dot{x}_{i}^{\alpha}\equiv \frac{dx_{i}^{\alpha}}{dt}$.  The quantity ${{\bf f}}_{i}\left(t\right)$ represents the collective effects of the solvent molecules on the colloid, and it is expressed by a stochastic force over the $i$-th particle, which satisfies the fluctuation-dissipation theorem in the Euclidean space $\mathbb{R}^{d+1}$, that is, $\left<{{\bf f}}_{i}\left(t\right)\right>=0$ and $\left<{{\bf f}}_{i}(t){{\bf f}}_{j}(\tau)\right>=2\zeta k_{B}T {\bf 1} \delta_{ij}\delta\left(t-\tau\right)$, where $k_{B}T$ is the thermal energy being $T$ the temperature and $k_{B}$ the Boltzmann's constant. The term ${\bf F}_{ij}\left(x_{i}, x_{j}\right)$ is the force that the $i$-th particle experiences at the position $x_{i}$ and is due to the interaction with the $j$-th particle located at the position $x_{j}$. In Eq. (\ref{many-particle-langevin}), the tangent vector ${\bf e}_\alpha\equiv \partial_{\alpha}{\bf X}$ projects the dynamics on the tangent space $T_{X}\left(\mathbb{M}\right)$, since the dynamics is occurring intrinsically on the manifold. Note that rising and lowering indices are done by the standard fashion using the metric tensor to lowering indices and inverse metric tensor $g^{\alpha\beta}$ for rising indices, for instance, $v^{\alpha}=g^{\alpha\beta}v_{\beta}$ for certain vector $v$. 

In the present exposition, we adopt the consideration that Eq. (\ref{many-particle-langevin}) is a set of $N$ stochastic differential equations in the Stratonovich's sense \cite{gardiner2009},
\begin{equation}
 dx_i^\alpha=\frac{1}{\zeta} \mathcal{F}^{\alpha}_{i} dt+\sqrt{2D_0}e_{i,a}^\alpha dW_{i,a}(t),
\label{langevin_ede_strat}
\end{equation}
where $\mathcal{F}^{\alpha}_{i}\equiv\sum_{j\neq i}F_{ij}^\alpha$, with $F_{ij}^\alpha$ as the tangent projection of the interacting term ${{\bf F}}_{ij}$,  and $D_{0}=k_{B}T/\zeta$ is the self-diffusion coefficient. Also, there is an implicit sum over the   indices $a=1, \cdots, d+1$, to take into account the tangent projection with the   
stochastic term in Eq. (\ref{many-particle-langevin}), which has been identified with a Wiener process for each particle $d{\bf W}_i(t)=(dW_{i,1}(t),dW_{i,2}(t),\cdots, dW_{i,d+1}(t))$, so that the total Wiener process $d{\bf W}(t)$ is such that $\text{dim}[d{\bf W}(t)]=(d+1)N$. Since the dynamics occurs on the curved space, the Wiener process should also be projected on it. So, we introduce a block diagonal projection operator ${\bf \hat{P}}=\text{diag}(e_{1, a}^\alpha,e_{2, a}^\alpha,...,e_{N, a}^\alpha)$, with $e_{i, a}^{\alpha}\equiv\left({\bf e}^{\alpha}\left(x_{i}\right)\right)_{a}$, where the blocks are individual operators for each particle given by the tensorial product of the basis of the tangent space and the basis of the Euclidean space. 
It is a well-known fact that given a differential stochastic equation in the Stratonovich form such as Eq. (\ref{langevin_ede_strat}), one can find its associate Chapman-Kolmogorov differential equation for the joint probability density function $p:\mathbb{M}^{N}\times \mathbb{R}\to \mathbb{R}$ \cite{gardiner2009}. For this, we only have to identify the components of the drift vector, and the diffusion matrix, which in this case are, $A_i^\alpha=\mathcal{F}^{\alpha}_{i}/\zeta $, and $B_{i,a}^\alpha=\sqrt{2D_0}e_{i, a}^\alpha$, respectively. Then, we obtain the following expression,
\begin{equation}
\partial_t p= -\frac{1}{\zeta} \sum_{i=1}^N\partial_\alpha\left (\mathcal{F}^{\alpha}_{i} p\right) + D_0\sum_{i=1}^N \partial_\alpha\left [e_{i, a}^\alpha \partial_\beta \left (e_{i, a}^\beta p\right )\right ].
\label{Eq3}
\end{equation}
In this equation, let us note that the partial derivation $\partial_{\alpha}=\frac{\partial}{\partial x^{\alpha}_{i}}$ depends on the index $i$, which is associated to the particle label. Although this last equation  has information on the geometry of the surface through the tangent vectors, it is not written in a covariant form yet. To this end,  we define the probability density  appropriately normalized with the volume element $dV=\prod_{i=1}^{N}dv^{i}_{g}$, where $dv^{i}_{g}$ is the Riemannian volume element defined by $dv^{i}_{g}\equiv d^{d}x_{i}\sqrt{g(x_{i})}$ for each particle. Thus, it is convenient to define a covariant joint probability density function $\rho\left(x_{1}, \cdots, x_{N};t\right)$ such as $p\left(x_{1}, \cdots, x_{N}; t\right)=\left(\prod_{i=1}^{N}\sqrt{g(x_{i})}\right)\rho\left(x_{1}, \cdots, x_{N};t\right)$, where $g(x_{i})$ is the determinant of the metric tensor $g_{\alpha\beta}\left(x_{i}\right)$. After this change and  using the Weingarten-Gauss equation mentioned above, Eq. (\ref{Eq3}) takes the following mathematical form,
\begin{equation*}
    \begin{split}
        \partial_t \rho
        &=-\frac{1}{\zeta} \sum_{i=1}^N  \nabla_{\alpha, i} \left (\mathcal{F}^{\alpha}_{i} \rho\right )+ D_0 \sum_{i=1}^N \frac{1}{\sqrt{g}} \partial_\alpha \left[g^{\alpha \beta}\sqrt{g}\partial_\beta \rho+g^{\alpha\beta}\rho(\partial_\beta \sqrt{g}-\sqrt{g}\Gamma_{\nu\beta}^\nu)\right],
    \end{split}
\end{equation*}
where the covariant derivative acting on a vector field $v^{\alpha}$ is $\nabla_{\alpha, i} v^{\alpha}=\frac{1}{\sqrt{g}}\partial_{\alpha}\left(\sqrt{g}\;v^{\alpha}\right)$ using the coordinates of the $i-$th particle, $x^{\alpha}_{i}$, and $\Gamma_{\nu\beta}^\alpha$ are the Christoffel symbols \cite{NakaharaBook}. Additionally, applying the identity $\Gamma_{\nu\beta}^\nu=\partial_\beta \log\sqrt{g}$, and identifying that the Laplace-Beltrami operator acts on scalars, $\Delta_{g, i}=\left(\sqrt{g}\right)^{-1}\partial_\alpha(g^{\alpha \beta}\sqrt{g}\partial_\beta)$ (also using local coordinates, $x^{\alpha}_{i}$), it is straightforward to obtain the desired covariant expression,
\begin{equation}
    \partial_t \rho= D_0\sum_{i=1}^N\Delta_{g, i} \rho -\frac{1}{\zeta} \sum_{i=1}^N\nabla_{\alpha, i}\left (\mathcal{F}^{\alpha}_{i}\rho\right ).
    \label{covariant_SE}
\end{equation}
\begin{figure}
\centering
\includegraphics[width=0.42\textwidth]{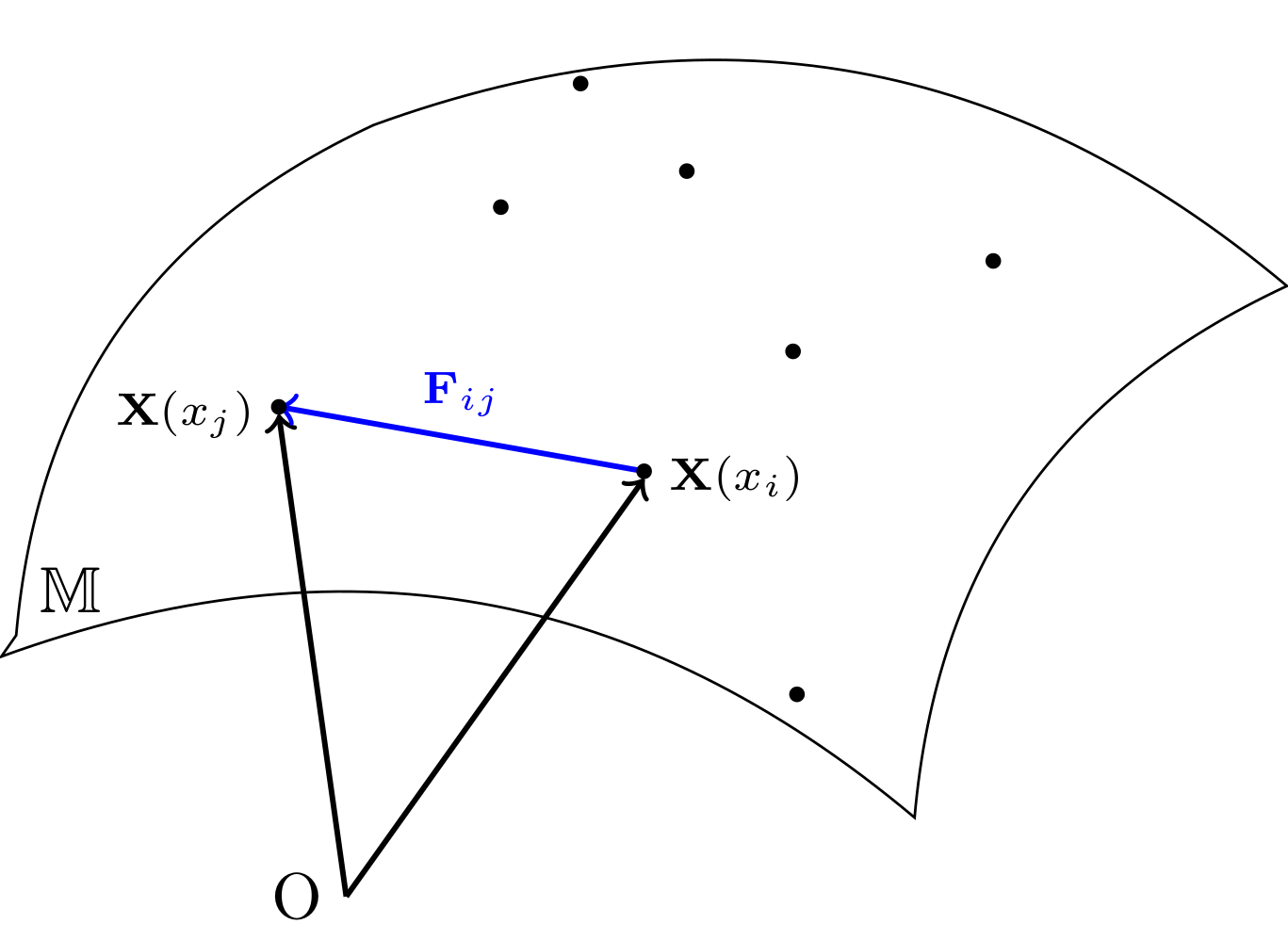} ~~~~~~~~~~~~~~~~~~\includegraphics[width=.32\textwidth]{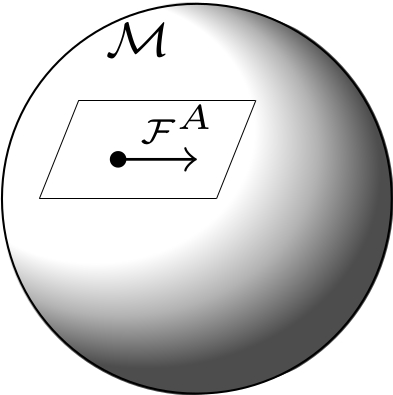}   
\caption{Left: Schematic representation of a set of particles embedded a manifold $\mathbb{M}$ of dimension $d$. The position of the particles is given by the embedding function ${\bf X}(x_i)$, and the force of interaction depends on the Euclidean distance measured in $\mathbb{R}^{d+1}$. Right: Schematic representation of a single particle in the manifold $\mathcal{M}=\mathbb{M}^{N}$. The particle is carried by an external force given by vector field $\boldsymbol{\mathcal{F}}^A$. Although both situations, left and right, seem to represent different systems, they are exactly the same physical problem.} 
\label{fig1}
\end{figure}
Equation (\ref{covariant_SE}) represents the covariant formulation of the Smoluchowski equation of a colloidal system of interacting particles constrained to a curved space $\mathbb{M}$, where all the geometrical features are included in the Laplace-Beltrami operator and the covariant derivative. This equation is reduced to the standard Smoluchowski equation when the manifold $\mathbb{M}$ is the open Euclidean space $\mathbb{R}^{d}$, where the metric tensor is $g_{\alpha\beta}=\delta_{\alpha\beta}$. 

Furthermore, one can write down equation (\ref{covariant_SE}) in a more compact form that allows us to prove that both systems shown in figure (\ref{fig1}), that is, the system of $N$ interacting particles confined to a $d$-dimensional manifold $\mathbb{M}$, and the system of a single particle in an external force confined to a $\mathcal{D}$-dimensional manifold $\mathcal{M}$ represent equivalent systems. 
For this purpose, let us define a hyper-dimensional Riemannian geometry by $N$ cartesian products of the manifold $\mathbb{M}$, that is,  $\mathcal{M}=\mathbb{M}\times \mathbb{M}\times \cdots\times \mathbb{M}\equiv\mathbb{M}^{N}$ of dimension $\mathcal{D}=N d$, where a local patch is described with the local coordinates $\xi^{A}=\left\{x^{\alpha}_{i}\right\}$, where we run the local indices $\alpha$ and the particles indices $i$, and $A=1, \cdots, \mathcal{D}$. Now, this manifold $\mathcal{M}$ is equipped with a  Riemannian metric defined through the following line element,
  \begin{eqnarray}
 ds^{2}=\sum_{i=1}^{N}g_{\alpha\beta}\left(x_{i}\right)dx^{\alpha}_{i}dx^{\beta}_{i}, \label{lineelement}
 \end{eqnarray}
 in terms of the metric tensor $g_{\alpha\beta}$ of the coordinates of each particle. Thus, the metric tensor associated with the line element (\ref{lineelement}) for the manifold $\mathcal{M}$ is given by the block diagonal matrix $G_{AB}={\rm diag}\left(g_{\alpha\beta}\left(x_{1}\right), \cdots, g_{\mu\nu}(x_{N})\right)$. It is not difficult to see that the covariant derivative compatible with the metric $G_{AB}$ for the manifold $\mathcal{M}$ can be written as,
 \begin{eqnarray}
 \boldsymbol{\nabla}_{A}=\left(\nabla_{\alpha, 1}, \nabla_{\beta, 2}, \cdots, \nabla_{\mu. N}\right),
 \end{eqnarray}
and the corresponding Laplace-Beltrami acting on scalars is simply the sum of each Laplace-Beltrami operator,
 \begin{eqnarray}
\boldsymbol{\Delta}_{G}=\boldsymbol{\nabla}_{A}\boldsymbol{\nabla}^{A}=\sum_{i=1}^{N}\Delta_{g, i}. 
 \end{eqnarray}
Now, defining $\boldsymbol{\mathcal{F}}^{A}=\left(\mathcal{F}^{\alpha}_{1}, \mathcal{F}^{\beta}_{2}, \cdots,\mathcal{F}^{\mu}_{N}\right)$ as the components of a vector field at the point $\xi\in \mathcal{M}$, it is straightforward to write down the Smoluchowski equation for the full $N-$particle colloidal system confined on the curved space (\ref{covariant_SE}) as,
\begin{equation}
    \partial_t \rho= D_0\boldsymbol{\Delta}_{G} \rho -\frac{1}{\zeta} \boldsymbol{\nabla}_{A }\left (\boldsymbol{\mathcal{F}}^{A}\rho\right ).
    \label{compactcovariant_SE}
\end{equation}
%

By expressing the Smoluchowski equation in this compact manner, it is now clear in what sense  one can interpret the problem of the interacting colloidal system as the Brownian motion of a single particle in an external field $\boldsymbol{\mathcal{F}}$ but in a hyper-dimensional space $\mathcal{M}$. This identification was already implemented in a previous contribution \cite{Castro-Villarreal_2021}, where an interacting colloidal system was studied on the line. Moreover, if we suppose that the interaction forces encoded in  $\boldsymbol{\mathcal{F}}^{A}$ can be written as $\boldsymbol{\mathcal{F}}_{A}=-\nabla_{A}\Phi$, where $\Phi$ is certain interacting potential, one can see that the expected equilibrium distribution is satisfied at long times, namely, $ \rho\left(\xi, t\right)=\frac{1}{\mathcal{Z}}e^{-\beta \Phi\left(\xi\right)}$, where $\mathcal{Z}$ is the partition function for the particle system confined to the curved manifold, 
\begin{eqnarray}
    \mathcal{Z}=\int \left(\prod_{i=1}^{N}dv^{i}_{g}\right)e^{-\beta \Phi\left(\xi\right)},
\end{eqnarray}
where $\beta=1/(\zeta D_{0})=1/(k_{B}T)$. Let us note that the expression of this partition function can also be obtained by integrating out the momentum $p^{\alpha}_{i}$ variables from the Boltzmann weight using the Hamiltonian $\mathcal{H}=\sum_{i=1}^{N}\frac{1}{2m}p^{\alpha}_{i}g_{\alpha\beta}(x_{i})p^{\beta}_{i}+\Phi\left(\xi\right)$. Usually, the potential $\Phi\left(\xi\right)$ is considered pairwise additive; thus, one can carry on the usual cluster diagrammatic expansion for the colloidal system in the curved space in static conditions \cite{tarjus2010}.


Consequently, equations (\ref{covariant_SE}) and (\ref{compactcovariant_SE}) represent the starting point of a covariant description that allows us to study in detail the colloid dynamics in curved spaces. In the following paragraphs, we will discuss a simple application of this formulation, and highlight some challenges and future perspectives that can be tackled within this approach.\\


\section{\LARGE Application of the covariant formulation: general behavior of the short-time dynamics in a dilute colloidal system}\label{Sect_III}

In this section, we study an application of the advantage to writing down the Smoluchowski equation in curved spaces in a covariant formulation (\ref{compactcovariant_SE}). This consists in providing a general behavior of the probability density function $\rho\left(\xi, \xi^{\prime}, t\right)$ at the short-time regime, or equivalently in a neighborhood around a point of the manifold $\mathcal{M}$. 
 Since $\mathcal{M}$ is a Riemannian manifold with metric tensor $G_{AB}$, one can explore the curvature effects on the colloidal interacting system using the Riemann normal coordinates (RNC), see, e.g., Refs. \cite{hatzinikitas2000note, NakaharaBook}, in the neighborhood of a point $p\in \mathcal{M}$, in an entirely analog manner as it has been performed for a single particle \cite{Castro2010}. 
 To derive an approximate expression for the density probability function  (PDF) $\rho\left(\xi, \xi^{\prime}, t\right)$ at a short time, it is common to write the Smoluchowski equation (\ref{compactcovariant_SE}) as a heat-kernel equation,
\begin{eqnarray}
\left(\partial_{t}+\hat{\mathcal{O}}\right)\rho\left(\xi, \xi^{\prime}, t\right)=\frac{1}{\sqrt{G}}\delta\left(\xi-\xi^{\prime}\right)\delta\left(t\right),\label{Heat-Kernel}
\end{eqnarray}
where the operator $\hat{\mathcal{O}}$ is defined as $\hat{\mathcal{O}}=-D_{0}\boldsymbol{\Delta}_{G}+\frac{1}{\zeta}\boldsymbol{\nabla}_{A}\left(\boldsymbol{\mathcal{F}}^{A}\cdot\right)$. At the initial condition, $t\to 0$, the PDF acquires the form of a Dirac delta: $\rho(\xi, \xi^{\prime}, t\to 0)=\frac{1}{\sqrt{G}}\delta\left(\xi-\xi^{\prime}\right)$. This initial condition establishes that the system is at the configuration $\xi^{\prime}$ at the starting time. Then, by performing a  Fourier transform on the time parameter, the above equation can be written as $\left(iE+\hat{\mathcal{O}}\right)\rho\left(\xi, \xi^{\prime}, E\right)=\delta\left(\xi-\xi^{\prime}\right)/\sqrt{G}$. 

In the following,  we use the De Witt procedure \cite{DeWittBook}; that is, we first separate the points to write the term $\sqrt{G}$ in front of the delta Dirac as the expression $\sqrt{G}\to G^{\frac{1}{4}}(\xi)G^{\frac{1}{4}}\left(\xi^{\prime}\right)$. Now, we redefine the PDF as $\overline{\rho}\left(\xi, \xi^{\prime}, t\right)=G^{\frac{1}{4}}\left(\xi\right)\rho\left(\xi, \xi^{\prime}, t\right)G^{\frac{1}{4}}\left(\xi^{\prime}\right)$. Thus, after some algebraic rearrangements, the above equation (\ref{Heat-Kernel}) can be rewritten as,
\begin{eqnarray}
    \left(iE+\hat{H}\right)\overline{\rho}\left(\xi, \xi^{\prime}, E\right)=\delta\left(\xi-\xi^{\prime}\right),\label{ResolventEq}
\end{eqnarray}
where $\hat{H}=G^{\frac{1}{4}}\hat{\mathcal{O}}G^{-\frac{1}{4}}$, or explicitly, this operator is given by 
\begin{eqnarray}
    \hat{H}=-D_{0}\left[\partial_{A}G^{AB}\partial_{B}+G^{-\frac{1}{4}}\partial_{A}\left(G^{\frac{1}{2}}G^{AB}\partial_{B}G^{-\frac{1}{4}}\right)-\beta\left(\partial_{A}\left(\boldsymbol{\mathcal{F}}^{A}\cdot\right)+\frac{1}{4}G^{-1}(\partial_{A}G)\boldsymbol{\mathcal{F}}^{A} \right)\right].\label{OpHam}
\end{eqnarray}
Next, we choose Riemann normal coordinates  $y^{A}$ in a local neighborhood $N_{\xi^{\prime}}\in \mathcal{M}$ centered at $\xi^{\prime}$. In RNC,  the neighborhood $N_{\xi^{\prime}}$ looks like Euclidean space, so we choose $\xi^{\prime}$ to be the origin of this Euclidean space.  The advantage of these coordinates is that one can express the metric tensor as $G_{AB}=\delta_{AB}+\frac{1}{3}\boldsymbol{\mathcal{R}}_{ACDB}~y^{C}y^{D}+\cdots$, where $\boldsymbol{\mathcal{R}}_{ACDB}$ is the Riemann curvature tensor of $\mathcal{M}$ evaluated at $\xi^{\prime}$.   In addition, we express the interaction terms in a  Taylor expansion around the origin of the neighborhood {\small $\boldsymbol{\mathcal{F}}^{A}\left(\xi\right)=\left(\boldsymbol{\mathcal{F}}^{A}\right)\left(\xi^{\prime}\right)+\left(\boldsymbol{\nabla}_{B}\boldsymbol{\mathcal{F}}^{A}\right)\left(\xi^{\prime}\right) y^{B}+\frac{1}{2}\left(\boldsymbol{\nabla}_{B}\boldsymbol{\nabla}_{C}\boldsymbol{\mathcal{F}}^{A}\right)\left(\xi^{\prime}\right)y^{B}y^{C}+\cdots$}, where the coefficients are evaluated at the point $\xi^{\prime}$. 

In the subsequent,  we have all the pieces to split the operator (\ref{OpHam})  as $\hat{H}=\hat{H}_{0}+\hat{H}_{I}$, where 
\begin{eqnarray}
    \hat{H}_{0}=D_{0}\hat{\bf p}^{2}-\frac{D_{0}}{6}\boldsymbol{\mathcal{R}}+D_{0}\beta \boldsymbol{\nabla}_{A}\boldsymbol{\mathcal{F}}^{A},
\end{eqnarray}
is a free ``Hamiltonian" and 
\begin{eqnarray}
    \hat{H}_{I}&=&D_{0}\beta\left(\boldsymbol{\nabla}_{B}\boldsymbol{\nabla}_{A}\boldsymbol{\mathcal{F}}^{A}-\frac{1}{6}\boldsymbol{\mathcal{R}}_{BA}\boldsymbol{\mathcal{F}}^{A}\right)y^{B}-\frac{D_{0}\beta}{6}\left(\boldsymbol{\mathcal{R}}_{BA}\boldsymbol{\nabla}_{C}\boldsymbol{\mathcal{F}}^{A}\right)y^{B}y^{C}+iD_{0}\beta \boldsymbol{\mathcal{F}}^{A}\hat{p}_{A}\nonumber\\
    &+&iD_{0}\beta \left(\boldsymbol{\nabla}_{B}\boldsymbol{\mathcal{F}}^{A}\right)y^{B}\hat{p}_{A}+i\frac{D_{0}\beta}{2}\left(\boldsymbol{\nabla}_{B}\boldsymbol{\nabla}_{C}\boldsymbol{\mathcal{F}}^{A}\right)y^{B}y^{C}\hat{p}_{A}-\frac{D_{0}}{3}\boldsymbol{\mathcal{R}}_{CABD}~\hat{p}^{A}y^{C}y^{D}\hat{p}^{B},
\end{eqnarray}
an interacting ``Hamiltonian", where we have defined a ``momentum operator" as $\hat{p}_{A}=-i\partial_{A}$ in analogy with quantum mechanics.  Now, the solution for the probability density function can be obtained by identifying $\delta(\xi-\xi^{\prime})=\left<\xi\right|\left.\xi^{\prime}\right>$ and solving equation (\ref{ResolventEq}) as follows $\overline{\rho}\left(\xi, \xi^{\prime}, E\right)= \left<\xi\left|\hat{K}\right|\xi^{\prime}\right> $,
where $\hat{K}=1/(iE+\hat{H})$ is the resolvent operator. Now, we carry on a standard perturbation theory at first order in an entire analogy with quantum mechanics; thus, the approximation of the resolvent operator through the perturbation theory is $\hat{K}=\hat{K}_{0}+\hat{K}_{0}\hat{H}_{I}\hat{K}_{0}+\cdots$. At this approximation, there are just $6$ terms to evaluate corresponding to quantities of the form $I_{i}\left(\xi, \xi^{\prime}\right)=\left<\xi\left|\hat{K}_{0}\hat{\mathcal{O}_{i}}\hat{K}_{0}\right|\xi^{\prime}\right>$, with $i=1, \cdots, 6$, where $\hat{\mathcal{O}}_{i}$ is one of the six terms: $y^{B}$, $y^{B}y^{C}$, $\hat{p}_{A}$, $y^{B}\hat{p}_{A}$, $y^{B}y^{C}\hat{p}_{A}$, and $\hat{p}^{A}y^{C}y^{D}\hat{p}^{B}$, respectively. Since $\hat{K}_{0}$ depends just on the ``momentum operator" $\hat{\bf p}$,  it is convenient to introduce two completeness relations using the momentum basis $\{\left|{\bf p}\right>\}$ to compute the contributions from the interacting Hamiltonian. Thus, one can write 
\begin{eqnarray}
I_{i}\left(\xi, \xi^{\prime}\right)=\int \frac{d^{\mathcal{D}}p}{\left(2\pi\right)^{\mathcal{D}}}\int d^{\mathcal{D}}q ~K_{0}\left(p, \alpha_{*}\right)e^{i\xi\cdot p}\left<{\bf p}\left|\hat{\mathcal{O}}_{i }\right|{\bf q}\right>K_{0}(q, \alpha_{*})e^{-i\xi^{\prime}\cdot q},\label{refint}
\end{eqnarray}
where $\alpha_{*}=-\frac{D_{0}}{6}\boldsymbol{\mathcal{R}}+D_{0}\beta \boldsymbol{\nabla}_{A}\boldsymbol{\mathcal{F}}^{A}$ and $K_{0}(p, \alpha)=1/(iE+D_{0}p^{2}+\alpha)$  is simply a function of value of the ``momentum" $p=\sqrt{p_{A}p^{A}}$ and energy $E$. In addition, we have used the transformation from the position to momentum basis as usual $\left<\xi\left|\right.{\bf p}\right>=e^{i\xi \cdot {p}}/(2\pi)^{\frac{\mathcal{D}}{2}}$. We should recall that we have chosen $\xi^{\prime}=0$ as the origin of the neighborhood $N_{\xi^{\prime}}$; this allows us to simplify the calculation of the integrals $I_{i}\left(\xi, \xi^{\prime}\right)$. In the appendix, we explicitly explain the procedure implemented to evaluate these integrals. After a straightforward calculation,  the short-time approximation for the probability density function $\rho(\xi, 0, t)$ of the full interacting system can be written as,
\begin{eqnarray}
\sqrt{G}\rho(\xi, 0, t)&=&\frac{1}{\left(4\pi D_{0}t\right)^{\mathcal{D}/2}}e^{-\frac{\xi^2}{4D_{0}t}}\left\{1+\tau^{(0)}+\tau^{(1)}_{B}\xi^{B}+\tau^{(2)}_{BC}\xi^{B}\xi^{C}+\cdots\right\},\label{STpdf}
\end{eqnarray}
where the terms $\tau^{(0)}$, $\tau^{(1)}_{B}$, $\tau^{(2)}_{BC}$ are tensors given by
\begin{eqnarray}
\tau^{(0)}&=&\left(D_{0}t\right)\left[\frac{1}{6}\boldsymbol{\mathcal{R}}-\frac{1}{2}\beta\boldsymbol{\nabla}_{A}\boldsymbol{\mathcal{F}}^{A}\right],\\
\tau^{(1)}_{B}&=&\frac{\beta}{2}\left[G_{BA}\left(1+D_{0}t\left(\frac{\boldsymbol{\mathcal{R}}}{6}-\beta\boldsymbol{\nabla}_{C}\boldsymbol{\mathcal{F}}^{C}\right)\right)+\frac{D_{0}t}{6}\left(\boldsymbol{\mathcal{R}}_{BA}+G_{BA}\boldsymbol{\nabla}_{G}-16\boldsymbol{\nabla}_{B}\boldsymbol{\nabla}_{A}\right)\right]\boldsymbol{\mathcal{F}}^{A},\\
\tau^{(2)}_{BC}&=&\frac{\beta}{4}\left[\left(1+D_{0}t\left(\frac{\boldsymbol{\mathcal{R}}}{6}-\beta\boldsymbol{\nabla}_{A}\boldsymbol{\mathcal{F}}^{A}\right)\right)\boldsymbol{\nabla}_{B}\boldsymbol{\mathcal{F}}_{C}-\frac{2D_{0}t}{9}\boldsymbol{\mathcal{R}}_{BA}\boldsymbol{\nabla}_{C}\boldsymbol{\mathcal{F}}^{A}\right]\nonumber\\
&-&\frac{1}{12}\left(1+D_{0}t\left(\frac{\boldsymbol{\mathcal{R}}}{6}-\frac{1}{2}\beta\boldsymbol{\nabla}_{C}\boldsymbol{\mathcal{F}}^{C}\right)\right)\boldsymbol{\mathcal{R}}_{BC}.
\end{eqnarray}
Equation (\ref{STpdf}) represents the probability distribution function of the interacting particle system at the short-time regime\footnote{Note that one can show that $\rho(\xi, 0, t)$ is normalized order by order in the perturbation theory of powers of $(D_{0}t)^{2}$. Indeed, using the above expectation values, $\left<1\right>=1+\tau^{(0)}+2D_{0}t \tau^{(2)}$, where $\tau^{(2)}=G^{AB}\tau^{(2)}_{AB}$, thus at the first order $D_{0}t$, one has $\left<1\right>=1+\left(D_{0}t\right)\left[\frac{1}{6}\boldsymbol{\mathcal{R}}-\frac{1}{2}\beta\boldsymbol{\nabla}_{A}\boldsymbol{\mathcal{F}}^{A}\right]-\frac{1}{6}D_{0}t \boldsymbol{\mathcal{R}}+\frac{\beta}{2}D_{0}t\boldsymbol{\nabla}_{A}\boldsymbol{\mathcal{F}}^{A}\approx 1$. 
}; it can be appreciated that the leading term, {\small $\rho_{0}(\xi,0, t)\equiv\exp\left[-\xi^2/(4D_{0}t)\right]/\left(4\pi D_{0}t\right)^{\mathcal{D}/2}$}, is given by the  Gaussian probability density valid for a very dilute system, while the subleading terms capture the corrections due to the curvature effects and interactions. 

Expectation values of observables can be calculated  using the standard definition $\left<O(\xi)\right>=\int_{\mathcal{M}} d^\mathcal{D} \xi\sqrt{G}\rho(\xi, 0, t)O(\xi)$.  Within the above approximation (\ref{STpdf}), the expectation values can be estimated in the short time using expectation values $\left<O(\xi)\right>_{0}$ using the leading term $\rho_{0}(\xi,0, t)$; that is {\small $\left<O(\xi)\right>=\left<O(\xi)\right>_{0}\left(1+\tau^{(0)}\right)+\tau^{(1)}_{B}\left<\xi^{B}O(\xi)\right>_{0}+\tau^{(2)}_{BC}\left<\xi^{B}\xi^{C}O(\xi)\right>_{0}+\cdots$}. Expectation values of polynomial observables are particularly easy to compute due to the Gaussian structure of $\rho_{0}\left(\xi, 0, t\right)$.

Here, we are interested in the estimation of the mean-square geodesic displacement  $\left<s^{2}\right>$, where $s=\sqrt{\delta_{AB}\xi^{A}\xi^{B}}$ is the geodesic displacement in RNC \cite{hatzinikitas2000note}. Also, it is interesting to estimate the expectation value of the coordinate itself $\xi^{B}$. For these expectation values, 
it is not a very difficult task to show  by
the standard calculation of the moments of a Brownian motion in a $\mathcal{D}$ dimensional space that {\small $\left<1\right>_{0}=1$}, meaning the normalization of the leading distribution $\rho_{0}(\xi, 0, t)$, the vanish of the odd products {\small $\left<\xi^{A_{1}}\xi^{A_{2}}\cdots\xi^{A_{2k+1}}\right>_{0}=0$}, for any positive integer $k$,  and for even products {\small $\left<\xi^{B}\right>_{0}=\left<\xi^{B}\xi^2\right>_{0}=\left<\xi^{B}\xi^{C}\xi^A\right>_{0}=0$},  {\small $\left<\xi^A \xi^{B}\right>_{0}=2D_{0}t G^{AB}$}, and {\small $\left<\xi^A \xi^{B}\xi^{2}\right>_{0}=4\left(\mathcal{D}+2\right)\left(D_{0}t\right)^{2} G^{AB}$}, where $G^{BC}$ is evaluated at $\xi^{\prime}$. Since the above approximation neglects quadratic curvature effects that correspond to prefactors of the order of $\left(D_{0}t\right)^{3}$ in the mean-square displacement \cite{Castro2010}, we just present  the result up to order of $(D_{0}t)^{2}$. This means that we basically neglect the linear terms of $D_{0}t$ in $\tau^{(2)}_{BC}$; thus, the mean-square displacement for the full $N$-particle system is given by,
\begin{eqnarray}
    \left<s^2\right>=2\mathcal{D} D_{0}t -\left[\frac{2}{3}\boldsymbol{\mathcal{R}}-2\beta\boldsymbol{\nabla}_{A}\boldsymbol{\mathcal{F}}^{A}\right]\left(D_{0}t\right)^{2}+\cdots.
\end{eqnarray}
One can notice that in absence of the interaction term, that is, when $\boldsymbol{\mathcal{F}}^{A}=0$, the mean-square displacement reduces to the previous result at the order $(D_{0}t)^{2}$ \cite{Castro2010}. In addition, it is not difficult to elucidate that the subsequent correction of the order of $(D_{0}t)^{3}$ involves pre-factors where curvature and interactions are coupled, for instance,  the terms $\boldsymbol{\mathcal{R}}\boldsymbol{\nabla}_{A}\boldsymbol{\mathcal{F}}^{A}$ and $\boldsymbol{\mathcal{R}}_{BA}\boldsymbol{\nabla}_{C}\boldsymbol{\mathcal{F}}^{A}$ from the tensor $\tau^{(2)}_{BC}$ appear as pre-factors; the cubic correction will be computed elsewhere in a future communication. Also, note that similar terms appear in the expectation value of $\xi_{B}$, $\left<\xi_{B}\right>=2D_{0}t \tau^{(1)}_{B}$, explicitly
\begin{eqnarray}
    \left<\xi_{B}\right>= \beta D_{0}t \boldsymbol{\mathcal{F}}_{B}+\beta \left(D_{0}t\right)^{2}\left[G_{BA}\left(\frac{\boldsymbol{\mathcal{R}}}{6}-\beta\boldsymbol{\nabla}_{C}\boldsymbol{\mathcal{F}}^{C}\right)+\frac{1}{6}\left(\boldsymbol{\mathcal{R}}_{BA}+G_{BA}\boldsymbol{\nabla}_{G}-16\boldsymbol{\nabla}_{B}\boldsymbol{\nabla}_{A}\right)\right]\boldsymbol{\mathcal{F}}^{A}.   
\end{eqnarray}
One can notice that in absence of the curvature,  $\left<\xi_{B}\right>$ reduces to the well-known term $\beta D_{0}t \boldsymbol{\mathcal{F}}_{B}$, which establishes, on average, a preferential direction of the Brownian motion. Also, this equation shows how the curvature is coupled to the interaction term within the $(D_{0}t)^{2}$ approximation.  Finally, given an interacting force ${\bf F}_{ij}$ and specific manifold $\mathbb{M}$, one can compute the mean-square displacement for a tagged particle of the colloidal system by defining ${\rm MSD}(t)=\frac{1}{N}\left<\xi^{2}\right>$, which is a quantity calculated in several computer simulations \cite{Castaneda2021}. 



\section{\LARGE Colloid dynamics in non-Euclidean spaces: some challenges and perspectives}

The covariant form of the Smulochowski equation (\ref{covariant_SE}) opens up the possibility of developing a theoretical framework to study different interesting phenomena that cannot be understood with the standard Statistical Mechanics approximations based on a Euclidean formulation. For example, one of the topics that can be tackled with this approach is the initiation of spinodal separation of particles interacting with short-ranged attractive forces and constrained to curved space, in analogy with the procedure presented by Dhont in the case of a Euclidean space \cite{Dhont1996}. Following these ideas, we need to convert Eq. (\ref{covariant_SE}) into an expression for the probability density of one particle instead of the joint probability of all the particles. To this end, it is necessary to perform a hierarchy of equations that allows us to marginalize the joint probability. Once obtained the reduced Smoluchowski equation, it is necessary to take advantage of the short-range interactions to relate out-of-equilibrium phenomena with their counterparts in equilibrium. The connection between both cases, as usual, is made through approximations concerning the equilibrium values; at this point, there exists a wide range of ways to proceed. For instance, a perturbation approach can be combined with the Riemann normal coordinates formalism, Monge's parameterization, or covariant Fourier series to calculate all the relevant observables. On the other hand, a covariant Taylor expansion \cite{Avramidi2015} approach can also be performed to compare the results with their flat counterparts \cite{Dhont2015}. 

In addition, the covariant formalism provided by equation (\ref{covariant_SE}) can be straightforwardly employed to highlight the role of the geometry on the equilibrium equation of the state of colloidal dispersions embedded in a curved space, to elucidate the geometrical contributions during the onset of non-equilibrium states, such as gels and glasses, to study the dynamics of either passive or active colloidal particles on manifolds and to investigate the curvature affects on the structural, kinetic and phase transitions of attractive colloids, just to mention a few examples of interest in the Colloidal Soft Matter domain. As mere speculation and motivated by the recent contribution presented in Ref. \cite{Castro2023}, the formalism here presented can also be considered to study the dynamics of granular matter in curved manifolds.

Furthermore, the covariant compact form of the Smoluchowski equation (\ref{compactcovariant_SE}) allowed us to obtain an expression for the joint probability density function for the full system at the short-time regime. The method implemented can be extended to capture corrections of the order of $(D_{0}t)^{3}$. The short-time expression of the PDF can be used to give the curvature effects in the mean-square displacement and the search role of the coupling between the curvature and the interactions; for instance, using this procedure, we can choose specific interaction force and specific manifold $\mathbb{M}$ and give an estimation of the mean-square geodesic displacement  of a tagged particle of the colloidal system at short times. Moreover, the short-time expression of the PDF (\ref{STpdf}) can also serve to define a computational scheme to study the behavior of the full system using a modified Monte Carlo simulation that considers curvature effects. Additionally,  the covariant compact form (\ref{compactcovariant_SE}) allows us to formulate the $N$-particle system using a Feynman path integral representation following the steps already implemented in \cite{Castro-Villarreal_2021}.  Last but not least, the study of some limiting cases of equation (\ref{covariant_SE}) and (\ref{compactcovariant_SE}) will also serve as a benchmark to computational or molecular simulation schemes adapted to study the behavior of colloids in non-Euclidean spaces.


\section*{Author Contributions}
All authors contributed equally to this work.

\section*{Funding}
Authors acknowledge financial support from CONACyT (Grants Nos. 237425, 287067 and A1-S-9098), PRODEP (Grant No.~511-6/17-11852), and the University of Guanajuato (Grant No. 103/2023). 

\section*{\Large Acknowledgments}
The authors acknowledge interesting and stimulating scientific discussions with Dr. Alejandro Villada-Balbuena and Prof. Jos\'e M. M\'endez-Alcaraz.

\section*{\Large Appendix}
In this section, we present the calculations that allowed us to present the general behavior of the probability distribution function in the short-time regime discussed in section \ref{Sect_III}. It is not difficult to show that $I_{0}(\xi, 0):=\left<\xi\left|\hat{K}_{0}\right|0\right>=\mathcal{J}_{1}\left(\xi\right)$ and the first five integrals (\ref{refint})  are given by
\begin{eqnarray}
    I_{1}\left(\xi, 0\right)&=&-\frac{1}{2}\xi^{B}\mathcal{J}_{2}\left(\xi\right),\\
    I_{2}\left(\xi, 0\right)&=&8D^2\partial_{B}\partial_{C}\mathcal{J}_{4}\left(\xi\right)+2D\delta_{BC}\mathcal{J}_{3}\left(\xi\right),
    \\
    I_{3}\left(\xi, 0\right)&=&-i\partial_{A}\mathcal{J}_{2}\left(\xi\right),\\
    I_{4}\left(\xi, 0\right)&=&\frac{i}{2}\xi^{B}\partial_{A}\mathcal{J}_{2}\left(\xi\right)-\frac{i}{2}\delta^{B}_{A}\mathcal{J}_{2}\left(\xi\right),\\
    I_{5}\left(\xi, 0\right)&=&\frac{i}{2}\left[\xi^{C}\delta^{B}_{A}+\xi^{B}\delta^{C}_{A}\right]\mathcal{J}_{2}\left(\xi\right)+i\partial_{A}I_{2}\left(\xi, 0\right),
\end{eqnarray}
where $\mathcal{J}_{n}=\left.\frac{(-1)^{n-1}}{\left(n-1\right)!}\frac{\partial^{n-1}}{\partial\alpha^{n-1}}\mathcal{J}_{1}\right|_{\alpha=\alpha*}$,  and 
\begin{eqnarray}
    \mathcal{J}_{1}({\bf y})=\int \frac{d^{\mathcal{D}}p}{\left(2\pi\right)^{\mathcal{D}}} K_{0}(p, \alpha) e^{{\bf y}\cdot {\bf p}}.
\end{eqnarray}
Also, we have not included $I_{6}\left(\xi, 0\right)$, since it is possible to prove that contracting with Riemann curvature tensor $\boldsymbol{\mathcal{R}}_{CABD}$ the resulting term vanishes as it is proved in the appendix of \cite{Castro2017}. Now, returning to the expression of the probability distribution function $\overline{\rho}(\xi, \xi^{\prime}, t)$, one can write as follows, 
\begin{eqnarray}
    \overline{\rho}(\xi,0, t)=\int_{-\infty}^{\infty}\frac{dE}{2\pi}e^{iE t}\left(\sum_{i=0}^{5}I_{i}\left(\xi,0\right)\right).
\end{eqnarray}
Now, each integration on $E$ can be carried out by promoting $E$ to a complex variable $z$. Thus, the $E$ integration can be carried on using Cauchy theorem in a complex variable after identifying the integral with  
\begin{eqnarray}
 \oint_{\gamma} \frac{dz}{2\pi i}\frac{e^{zt}}{z+A}=e^{-At },   
\end{eqnarray}
where  $\gamma$ is a  clockwise semi-circular contour in the left half complex plane enclosed the pole $z_{0}=-A$ on the negative real line.   Now, by using the approximation for $G^{\frac{1}{4}}\left(\xi\right)\simeq 1-\frac{1}{12}\boldsymbol{\mathcal{R}}_{AB}\xi^{A}\xi^{B}$ in expression $\overline{\rho}(\xi, \xi^{\prime}, t)=G^{\frac{1}{4}}\left(\xi\right)\rho(\xi, \xi^{\prime}, t)G^{\frac{1}{4}}\left(\xi^{\prime}\right)$, one can get the final form of the probability density function $\rho(\xi, 0: t)$. Then, by putting all terms in powers of $\xi$, we finally get the short-time approximation given by Eq. (\ref{STpdf}). 



\bibliographystyle{plain} 

\bibliography{GeometryCSMP}

\begin{thebibliography}{10}

\bibitem{ALENGHAT2013}
Francis~J. Alenghat and David~E. Golan.
\newblock Chapter three - membrane protein dynamics and functional implications
  in mammalian cells.
\newblock In Vann Bennett, editor, {\em Functional Organization of Vertebrate
  Plasma Membrane}, volume~72 of {\em Current Topics in Membranes}, pages
  89--120. Academic Press, 2013.

\bibitem{ando2010}
Tadashi Ando and Jeffrey Skolnick.
\newblock Crowding and hydrodynamic interactions likely dominate in vivo
  macromolecular motion.
\newblock {\em Proceedings of the National Academy of Sciences},
  107(43):18457--18462, 2010.

\bibitem{Apaza2018}
Leonardo Apaza and Mario Sandoval.
\newblock Active matter on riemannian manifolds.
\newblock {\em Soft Matter}, 14:9928--9936, 2018.

\bibitem{Avramidi2015}
Ivan Avramidi.
\newblock {\em Heat Kernel Method and its Applications}.
\newblock 11 2015.

\bibitem{Basak2019}
Sujit Basak, Sengupta Sengupta, and Krishnananda Chattopadhyay.
\newblock Understanding biochemical processes in the presence of sub-diffusive
  behavior of biomolecules in solution and living cells.
\newblock {\em Biophysical Reviews}, 11:851--872, 2019.

\bibitem{Castaneda2021}
Ram\'on Casta{\~n}eda-Priego.
\newblock Colloidal soft matter physics.
\newblock {\em Revista Mexicana de F\'isica}, 67:050101, 2021.

\bibitem{Castro2010}
Pavel Castro-Villarreal.
\newblock Brownian motion meets riemann curvature.
\newblock {\em Journal of Statistical Mechanics: Theory and Experiment},
  2010(08):P08006, aug 2010.

\bibitem{Castro2014b}
Pavel Castro-Villarreal.
\newblock Intrinsic and extrinsic measurement for brownian motion.
\newblock {\em Journal of Statistical Mechanics: Theory and Experiment},
  2014(5):P05017, may 2014.

\bibitem{Castro-Villarreal_2021}
Pavel Castro-Villarreal, Claudio Contreras-Aburto, Sendic Estrada-Jiménez,
  Idrish Huet-Hernández, and Oscar Vázquez-Rodríguez.
\newblock Single file diffusion meets feynman path integral.
\newblock {\em Journal of Statistical Mechanics: Theory and Experiment},
  2021(9):093208, sep 2021.

\bibitem{Castro2017}
Pavel Castro-Villarreal and R.~Ruiz-S\'anchez.
\newblock Pseudomagnetic field in curved graphene.
\newblock {\em Phys. Rev. B}, 95:125432, Mar 2017.

\bibitem{Castro2018}
Pavel Castro-Villarreal and Francisco~J. Sevilla.
\newblock Active motion on curved surfaces.
\newblock {\em Phys. Rev. E}, 97:052605, May 2018.

\bibitem{Castro2014}
Pavel Castro-Villarreal, Alejandro Villada-Balbuena, José~Miguel
  Méndez-Alcaraz, Ramón Castañeda-Priego, and Sendic Estrada-Jiménez.
\newblock A brownian dynamics algorithm for colloids in curved manifolds.
\newblock {\em The Journal of Chemical Physics}, 140(21):214115, 2014.

\bibitem{DeWittBook}
Bryce~S DeWitt.
\newblock {\em Dynamical theory of groups and fields}.
\newblock Documents on modern physics. Gordon and Breach, no earlier edition
  stated edition, 1965.

\bibitem{Dhont1996}
Jan~K.G. Dhont.
\newblock {\em An Introduction to Dynamics of Colloids}.
\newblock ISSN. Elsevier Science, 1996.

\bibitem{Dhont2015}
Jan~K.G. Dhont.
\newblock {\em {D}ynamics of {C}olloids}, volume~94 of {\em Reihe
  Schlüsseltechnologien}, pages B1; 1--40.
\newblock Schriften des FZ-Jülich, Jülich, Feb 2015.

\bibitem{Einstein}
A.~Einstein.
\newblock Zur theorie der brownschen bewegung.
\newblock {\em Annalen der Physik}, 324(2):371--381, 1906.

\bibitem{Faraudo02}
Jordi Faraudo.
\newblock Diffusion equation on curved surfaces. i. theory and application to
  biological membranes.
\newblock {\em The Journal of Chemical Physics}, 116(13):5831--5841, 2002.

\bibitem{gardiner2009}
C.~Gardiner.
\newblock {\em Stochastic Methods: A Handbook for the Natural and Social
  Sciences}.
\newblock Springer Series in Synergetics. Springer Berlin Heidelberg, 2009.

\bibitem{hatzinikitas2000note}
Agapitos Hatzinikitas.
\newblock A note on riemann normal coordinates, 2000.

\bibitem{Holyst}
R.~Ho\l{}yst, D.~Plewczy\ifmmode~\acute{n}\else \'{n}\fi{}ski, A.~Aksimentiev,
  and K.~Burdzy.
\newblock Diffusion on curved, periodic surfaces.
\newblock {\em Phys. Rev. E}, 60:302--307, Jul 1999.

\bibitem{kafri2005}
Yariv Kafri, David~K. Lubensky, and David~R. Nelson.
\newblock Dynamics of molecular motors with finite processivity on
  heterogeneous tracks.
\newblock {\em Phys. Rev. E}, 71:041906, Apr 2005.

\bibitem{Castro2023}
M.~Ledesma-Motolin\'{\i}a, J.~L. Carrillo-Estrada, A.~Escobar, F.~Donado, and
  Pavel Castro-Villarreal.
\newblock Magnetized granular particles running and tumbling on the circle
  ${S}^{1}$.
\newblock {\em Phys. Rev. E}, 107:024902, Feb 2023.

\bibitem{NakaharaBook}
Mikio Nakahara.
\newblock {\em Geometry, topology, and physics}.
\newblock Graduate student series in physics. Institute of Physics Publishing,
  2nd ed edition, 2003.

\bibitem{Quintana2018}
C.~Quintana and P.~González-Mozuelos.
\newblock Nanoparticles confined to a spherical surface in the presence of an
  external field: Interaction forces and induced microstructure.
\newblock {\em The Journal of Chemical Physics}, 148(23):234901, 2018.

\bibitem{Ramadurai2009}
Sivaramakrishnan Ramadurai, Andrea Holt, Victor Krasnikov, Geert van~den
  Bogaart, J.~Antoinette Killian, and Bert Poolman.
\newblock Lateral diffusion of membrane proteins.
\newblock {\em Journal of the American Chemical Society}, 131(35):12650--12656,
  2009.
\newblock PMID: 19673517.

\bibitem{Ramirez2017}
O.~A. Ramírez-Garza, J.~M. Méndez-Alcaraz, and P.~González-Mozuelos.
\newblock Structural and dynamic inhomogeneities induced by curvature gradients
  in elliptic colloidal halos of paramagnetic particles.
\newblock {\em The Journal of Chemical Physics}, 146(19):194903, 2017.

\bibitem{D0CP06474B}
O.~A. Ramírez-Garza, J.~M. Méndez-Alcaraz, and P.~González-Mozuelos.
\newblock Effects of the curvature gradient on the distribution and diffusion
  of colloids confined to surfaces.
\newblock {\em Phys. Chem. Chem. Phys.}, 23:8661--8672, 2021.

\bibitem{Sarmiento2016}
Erick Sarmiento-G\'omez, Jos\'e~Ram\'on Villanueva-Valencia, Salvador
  Herrera-Velarde, Jos\'e~Arturo Ruiz-Santoyo, Jes\'us Santana-Solano,
  Jos\'e~Luis Arauz-Lara, and Ram\'on Casta\~neda Priego.
\newblock Short-time dynamics of monomers and dimers in quasi-two-dimensional
  colloidal mixtures.
\newblock {\em Phys. Rev. E}, 94:012608, Jul 2016.

\bibitem{tarjus2010}
Gilles Tarjus, Francois Sausset, and Pascal Viot.
\newblock Statistical mechanics of liquids and fluids in curved space, 2010.

\bibitem{villada2021}
Alejandro Villada-Balbuena, Antonio Ortiz-Ambriz, Pavel Castro-Villarreal,
  Pietro Tierno, Ram\'on Casta\~neda Priego, and Jos\'e~Miguel
  M\'endez-Alcaraz.
\newblock Single-file dynamics of colloids in circular channels: Time scales,
  scaling laws and their universality.
\newblock {\em Phys. Rev. Res.}, 3:033246, Sep 2021.

\bibitem{Villanueva2019}
Jos\'e~Ram\'on Villanueva-Valencia, Jes\'us Santana-Solano, Erick
  Sarmiento-G\'omez, Salvador Herrera-Velarde, Jos\'e~Luis Arauz-Lara, and
  Ram\'on Casta\~neda Priego.
\newblock Long-time dynamics and hydrodynamic correlations in
  quasi-two-dimensional anisotropic colloidal mixtures.
\newblock {\em Phys. Rev. E}, 98:062605, Dec 2018.

\bibitem{Zhong2017}
Yaning Zhong, Luyang Zhao, Paul~M. Tyrlik, and Gufeng Wang.
\newblock Investigating diffusing on highly curved water–oil interface using
  three-dimensional single particle tracking.
\newblock {\em The Journal of Physical Chemistry C}, 121(14):8023--8032, 2017.

\end{thebibliography}

\end{document}